\documentclass[aps,preprint,showpacs,showkeys,amsmath,amssymb,superscriptaddress,groupedaddress,showkeys]{revtex4}  
\usepackage{graphicx}  
\usepackage{epstopdf}
\usepackage{subfigure}
\usepackage{dcolumn}   
\usepackage{bm}        
\usepackage{amssymb}   
\usepackage{multirow}
\usepackage{indentfirst}
\usepackage{subfigure}
\usepackage{array}
\usepackage{xcolor}
\usepackage{rotating}
\usepackage{array,longtable}
\usepackage{appendix}

\newcommand{\Tsinghua}{\affiliation{Department~of~Engineering~Physics, Tsinghua~University, Beijing 100084, China}}
\newcommand{\Steel}   {\affiliation{China~Iron~\&~Steel~Research~Institute~Group, Beijing 100081, China}}

\begin{document}

\title{ Assay of low-background stainless steel by smelting for the neutrino experiment at Jinping}

\author{Ghulam Hussain\footnote{Corresponding author: gu-lm11@mails.tsinghua.edu.cn}}
\author{Zhi Zeng}\Tsinghua
\author{Chunfa Yao}\Steel
\author{Mohan Li}\Tsinghua
\author{Ziyi Guo}\Tsinghua
\author{Lei Guo}\Tsinghua
\author{Zhe Wang\footnote{Corresponding author: wangzhe-hep@mail.tsinghua.edu.cn}}
\author{Shaomin Chen}\Tsinghua

\date{\today}

\begin{abstract}
To ensure compliance with the experimental requirement for ultra-low background, in this study the radioactivity of stainless steels manufactured by smelting is thoroughly investigated.
Raw materials, stage samples, and commercial samples are investigated by glow discharge mass spectrometry (GDMS) and/or with high-purity germanium detectors (HPGe) at both the ground level and/or the China Jinping Underground Laboratory.
Custom-made stainless steel samples are found to have radioactivity levels comparable to those in other low-background experiments.
The comprehensive results regarding  the radioactivity level in materials to be used in the proposed Jinping Neutrino Experiment are reported.
\end{abstract}


\maketitle

\section{Introduction}
Due to its good mechanical strength, cost effectiveness, ease of processing, and stability in the presence of many chemicals, stainless steel (SST) is a favorite material for use in particle experiments.
To use SST in low-background experiments, such as  neutrino or dark matter experiments,
the radioactivity of SST must be determined.
The Borexino~\cite{Borexino-2002}, Gerda~\cite{Gerda-2008}, NEXT~\cite{Next-2015}, PandaX~\cite{PandaX-2016,PandaX-2016-2} experiment, etc.~set a rather high standard for their low-background SST.
In this paper, we investigate the radioactivity of SST produced by smelting,
as the radioactivity of $\rm ^{238}U$, $\rm ^{232}Th$, $\rm ^{40}K$, and $\rm ^{60}Co$ must all  be at the level of a few mBq/kg to meet the requirements of the Jinping Neutrino Experiment~\cite{Jinping, Jinping-Geo, Jinping-SRN}.

Since SST will be used to construct the detector framework of the Jinping Neutrino Experiment,
it will have direct contact with the water or liquid scintillator. SST grades 304L and 316L are of greatest interest for their corrosion-resistance property.
The smelting process is often used to produce new SST samples for laboratory research purposes.
It uses pure material as input, and is thought to be easier to control contaminations.
Adequate facilities are available to produce a reasonable amount of SST for  particle experiments.
This process is investigated here to understand how radioactivity is introduced.

In this article, we analyzed the samples produced by the China~Iron~\&~Steel~Research~Institute~Group (CISRIG).
In Section~\ref{SMP}, the smelting process is explained in detail, and the samples to be tested are introduced.
In Section~\ref{AM}, the assay methods for the low background counting are described.
Section~\ref{TPAR} presents the testing procedure and results.
In Section~\ref{CR}, the results are compared with those of other experiments.
Lastly, conclusions are given in section ~\ref{CONC}.

\section{Smelting process and samples}\label{SMP}
\subsection{Stainless steel}
Grade 304 SST is an austenitic grade that can be severely deep drawn,
and this feature makes 304 the dominant grade for use in applications such as pans and sinks.
It also has good corrosion resistance in many environments and when in contact with corrosive media.
Grade 304L SST is the low-carbon version of 304, which yields improved weldability.

 Grade 316 SST is the standard molybdenum-bearing grade of 304. The molybdenum provides
better corrosion resistance than in grade 304, especially with respect to pitting and crevice corrosion in chloride environments.
Type 316L, the low-carbon version, is widely used in heavy-gauge welded components.

Table~\ref{tab:composition} shows the chemical formula of SST 304L and 316L, in which Fe, Cr, Ni are the basic composition elements for 304L and Mo is added to 316L.
For most of the commercially available SSTs, C, Si, P, S, and Mn are impurities,
and P and S in particular are harmful and should be removed. Other elements must also  be controlled to within certain limits, as shown in Table~\ref{tab:composition}.

\begin{table}[!h]
\begin{tabular}{cccccccccc}
\hline\hline
Element fraction (\%) &Fe  &Cr	  &Ni	 &C		      &Si		  &P            &S		    &Mn		    &Mo\\\hline
SST 304L        	  &68  &19.0  &10.0  &$\leq$0.03  &$\leq$1.0  &$\leq$0.035  &$\leq$0.03 &$\leq$2.0  &- \\
SST 316L              &64  &17	  &13.5  &$\leq$0.03  &$\leq$1.0  &$\leq$0.045  &$\leq$0.03 &$\leq$2.0  &2.5\\
\hline
\end{tabular}
\caption{Composition of the two types of stainless steel considered for peripheral structures.}
\label{tab:composition}
\end{table}

A typical neutrino experiment  should run for more  than ten years, and in some cases even more than 30 years.
They are also likely to take place underground to ensure  a large overburden as a shield for cosmic-ray muons.
Grades 304L and 316L  SST are the primary choices for the peripheral structures, which may have been in direct contact
with underground water or humid air.

\subsection{Smelting process}
Below, the smelting process used in producing the SST samples for this analysis is briefly described.
For simplicity, we focus only on the key points at which new materials are introduced or where possible contact with the environment is unavoidable. A more technical introduction to the smelting process can be found in reference~\cite{smelting}.
\begin{enumerate}
\item  A MgO sand crucible or mold is made and then placed inside a vacuum chamber. The smelting process is then performed in the MgO crucible, which can be maintained at a high temperature throughout the process.
\item  Cleaning the crucible: pure iron is imported into the pot and, while maintaining the pressure of the vacuum chamber, the pure iron is melted into liquid form. Then, the liquid iron is removed and loose surfaces of the crucible are washed.
\item  Pure Cr and Ni for 304L (and pure Mo for 316L) are imported into the crucible and, while maintaining the pressure of the chamber to a certain value, the pure iron and raw materials are melted together.
\item  To simulate commercial SST, small amounts of C, Si, P, S, and Mn can be introduced.
\item  When the liquid metal mixture containing all the raw materials is ready, the crucible and its liquid metal contents are cooled.
\item  The crucible mold and the outer ingot layer of the SST sample are then removed for the best mechanical property and the inner part of the ingot is used in the test and in other applications.
\end{enumerate}
In the smelting process, the input materials have purities of up to 99.7-99.9\%.
The types of all the crucial materials are ``YT01" for Fe, ``No. 1" for Ni, ``99A" for Cr, ``High purity" for Mo, and ``Class I" for MgO crucible sand.
During the whole process, the sample only has access to the MgO sand crucible.
The smelting process is simpler than the typical methods used in large SST factories to understand,
and thus contamination sources can be much better identified and controlled.

\subsection{Samples}
\subsubsection{Raw materials}
\begin{figure}
\includegraphics[width=15cm, height=15cm]{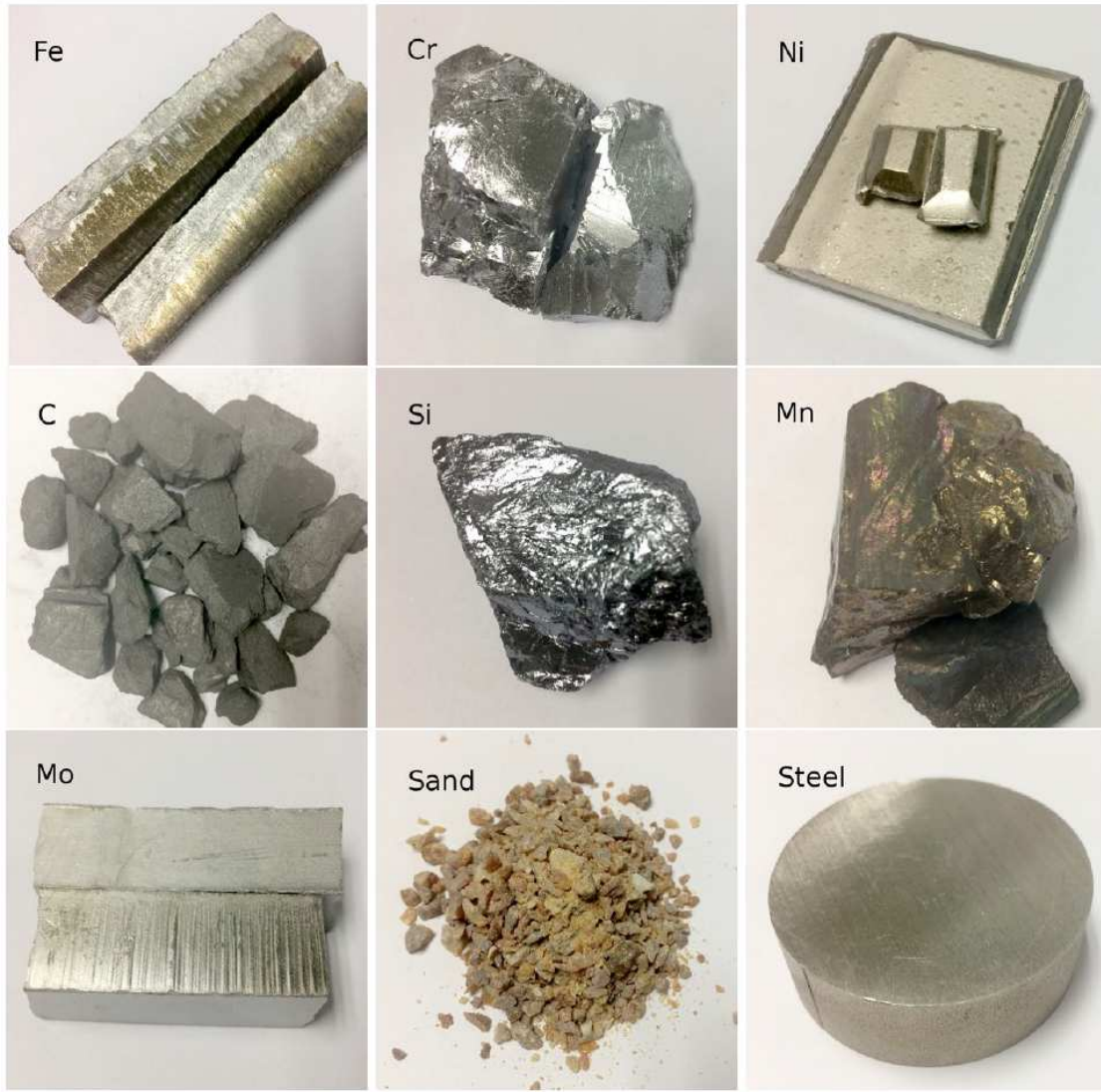}
\caption{Raw materials, MgO sand and a SST sample.}
\label{fig:rawsamp}
\end{figure}

Figure~\ref{fig:rawsamp} shows the raw materials of Fe, Cr, Ni, C, Si, Mn, and Mo. P and S were not considered.
The figure also shows photos of the MgO sand and one SST sample.

\subsubsection{Stage samples}
Given the smelting procedure used, the input materials are not the only potential sources of radioactive contamination.
The entire process may introduce radioactive impurities such as, for instance, those in the MgO sand.
Four stage samples are acquired for the test procedure.
\begin{enumerate}
\item The input Fe sample mentioned above, called ``Raw Iron".
\item  Iron ingot, which is the iron sample, that passed through the whole
smelting process without any other additions, called ``Ingot". This sample is used to test whether
a crucible made from sand can contaminate the iron sample, and whether there are any unknown
contamination processes.
\item SST 304L sample, called the ``CustomMade 304L".
\item SST 316L sample, called the ``CustomMade 316L". Note that in this and in the above CustomMade 304L  sample, no C, Si, P, S, and Mn were added, which will be explained later.
\end{enumerate}

\subsubsection{Commercial SST samples}
For comparison purposes, two commercial SST samples of 304L and 316L are acquired, called the ``Commercial 304L" and ``Commercial 316L".
Since there are many different SST suppliers and many different iron and other ores
can be used, our selected samples are not necessarily representative.

All samples were prepared with the same shape and dimensions for easy comparison. The mass of each sample is about 0.5 kg.

\section{Analysis methods}\label{AM}
\subsection{Glow discharge mass spectrometry}
Glow discharge mass spectrometry (GDMS) is a popular trace element measurement technique for inorganic solid materials.
The detection limit of GDMS can reach the ppt level, i.e., $1\times 10^{-9}$~g/g.
For this study the GDMS measurement was provided by the National Center of Analysis and Testing for Non-ferrous Metals and Electronics Materials, China. For most elements, the detection limit is $5\times10^{-9}$~g/g.
The measurement results in this study are semiquantitative. A precise quantitative measurement will further require several reference samples to calibration the response, which are not available for the specific application in this study.

\subsection{High-purity germanium detector at Jinping laboratory}
A low-background high-purity germanium gamma ray spectrometer was installed at the China Jinping Underground Laboratory (HPGe-Jinping).
The germanium detector is a coaxial n-type  germanium detector with 40\% relative detection efficiency and was fabricated by CANBERRA in France.
The germanium crystal is 59.9 mm in diameter and 59.8 mm in height.
 The detector is equipped with a cryostat made from ultra-low background aluminum.
As protection against the environmental gamma background and neutrons, from the inside out, the detector is shielded by 5 cm of oxygen-free-high-purity copper, 15 cm of lead, and 10 cm of borated polyethylene plate.
A preamplifier is located outside the shielding due to its associated relatively high radioactivity contamination.
The HPGe-Jinping has a large sample screening chamber with the dimensions  30 $\times$ 30 $\times$ 63 $\rm cm^{3}$.
During the data-taking period, the entire detector is flushed with boil-off nitrogen to reduce the $^{222}$Rn background in the air. The detection limits of $\rm ^{238}U$, $\rm ^{232}Th$, $^{226}$Ra, $\rm ^{40}K$, and $\rm ^{137}Cs$ are approximately 1~mBq/kg.
More details regarding the detector can be found in reference~\cite{Zeng-2014HPGe}.

\subsection{High-purity germanium detector at ground level}
Another high-purity germanium detector was installed in 2004 at ground level (HPGe-Ground) in the Department of Engineering Physics of Tsinghua University.
The HPGe-Ground is equipped with an HPGe spectrometer from ORTEC, an anti-Compton system, and a shielding system.
The anti-Compton system is made with NaI (Tl) crystal to veto Compton scattering events.
The shielding system uses lead and paraffin to suppress environmental gammas and neutrons.
A more detailed description of the HPGe-Ground can be found in reference~\cite{Zeng-2004}.
The detection limits of $\rm ^{238}U$, $\rm ^{232}Th$, $^{226}$Ra, $\rm ^{40}K$, and $\rm ^{137}Cs$ are approximately 1~Bq/kg limited by the detector background at surface level.

\section{Testing procedure and results}\label{TPAR}
\subsection{Raw materials inspection}
\subsubsection{Testing results}
The raw materials are cleaned with water and by supersonic cleaning whenever possible.
A portion of the carbon sample was in powder form and could not be washed, which was also the case for the powdered MgO sand, as shown in Fig.~\ref{fig:rawsamp}.

These materials were screened using the HPGe-Ground. The raw counting rate results indicated
the radioactivity of the MgO sand, C, and Si to be very high.
Of these, the counting rate of the MgO sand is the highest one, almost twice that of the background.
The counting rates of the C and Si samples are found to be slightly higher than that of the background.
The results for the other raw materials were compatible to the background and
the HPGe-Ground has reached its measurement limit for them.

\subsubsection{Conclusions and next step}
MgO sand must be used in the smelting process, but its high radioactivity is a matter of concern.
The radioactivity of samples must be determined before and after they come  into  contact with the MgO crucible, and stage samples were requested for this purpose.

C and Si samples had already exhibited  higher counting rates than the HPGe-Ground background,
and  were not added when designing the stage samples.
Since Mn is also not 100\% necessary and S and P are harmful, to avoid complexity, no C, Si, P, S, or Mn is added to
the CustomMade 304L and CustomMade 316L.

The sensitivity of the HPGe-Ground is low and we did not pursue more detailed quantified measurements.
\subsection{Inspection of stage and commercial samples}
\subsubsection{GDMS results}
\begin{figure}
\includegraphics[width=15cm]{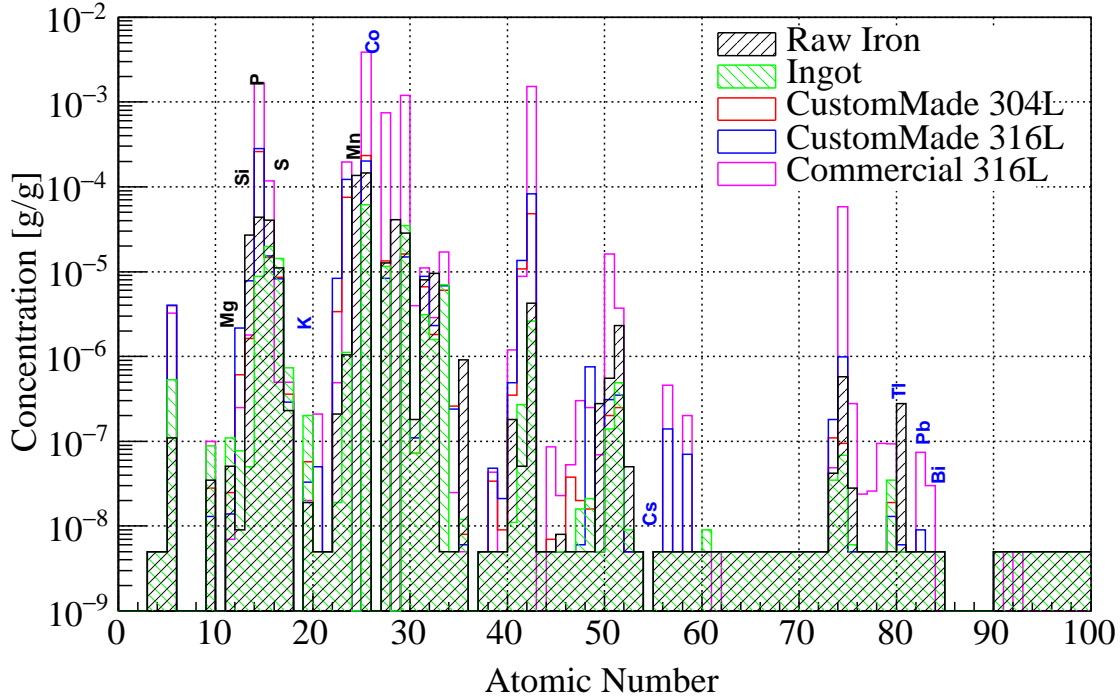}
\caption{ \label{fig:GDMS-profile}  Concentration of elements of Raw Iron, Ingot, CustomMade 304L, 316L and Commercial 316L sample by GDMS measurement. The complete results of these GDMS measurements are reported in Appendix~\ref{App}.
The harmful elements, Mg, Si, P, S, and Mn, for SST, and elements, K, Co, Cs, Tl, Pb, and Bi, with potential radioactive isotopes are labeled in black and blue, respectively.
}
\end{figure}

The element fractions of the Raw Iron, Ingot, CustomMade 304L and 316L, and Commercial 316L were analyzed by GDMS.
For comparison, Fig.~\ref{fig:GDMS-profile}  shows plots of fractions of all the elements in each sample.
The fractions of the harmful elements for SST, Mg, Si, P, S and Mn, and the elements with potential radioactive isotopes, K, Co, Cs, Tl, Pb, Bi, U, and Th of the CustomMade 316L and Commercial 316L samples are listed in Table~\ref{tab:GDMStable}.
Because the situation of Commercial 316L is much worse than CustomMade samples, we did not measure Commercial 304L with GDMS.
A full table of the results can be found in appendix~\ref{App}.

Significant improvement can be found in the CustomMade samples.
The fractions of P, Si, and Mn in the CustomMade 316L  sample are much lower than in the Commercial sample, although the S fraction is worse than that in the Commercial sample, and C cannot be measured by GDMS.
The elements with radioactive isotopes of concern, including Pb, Bi, Tl, K, Co, and Cs, are all better than or at a similar level to that of the Commercial sample.

From Raw Iron to Ingot to CustomMade 304L and 316L, the Mg element fraction was observed to increase by factors of 4-9 at each stage. The contamination of the MgO crucible was observed.

\begin{table}
 \footnotesize
\begin{tabular*}{73mm}{@{\extracolsep{\fill}}cccc}
\toprule Element         & Commercial 316L             & CustomMade 316L&        \\
&                        $ \times 10^{-9}$ g/g    &  $ \times 10^{-9}$g/g      \\\hline
U                        &$<$5              &$<$5          \\
Th                       &$<$5              &$<$5          \\
Pb                       &74                &9                 \\
Bi                       &30                &$<$5          \\
Tl                       &$<$5              &$<$5          \\
K                        &20                &33                       \\
Co                       &742$\times10^3$   &8390             \\
Cs                       &$<$5              &$<$5             \\
Mg                       &250               &2170 \\
S                        &$<$500            &8250              \\
P                        &118$\times10^3$    &15400             \\
Si                       &1671$\times10^3$    &284 $\times10^3$                         \\
Mn                       &3853$\times10^3$    &202     $\times10^3$                 \\
 \hline
\end{tabular*}
\caption{\label{tab:GDMStable}
GDMS measurement result of Commercial 316L and CustomMade 316L. Harmful elements to SST, S, P, Si, Mn, and Mg, and elements with radioactive isotopes Pb, Bi, Tl, K, Co, Cs, U, and Th are listed.}
\end{table}

\subsubsection{HPGe-Jinping results}

\begin{figure}
\includegraphics[width=15cm]{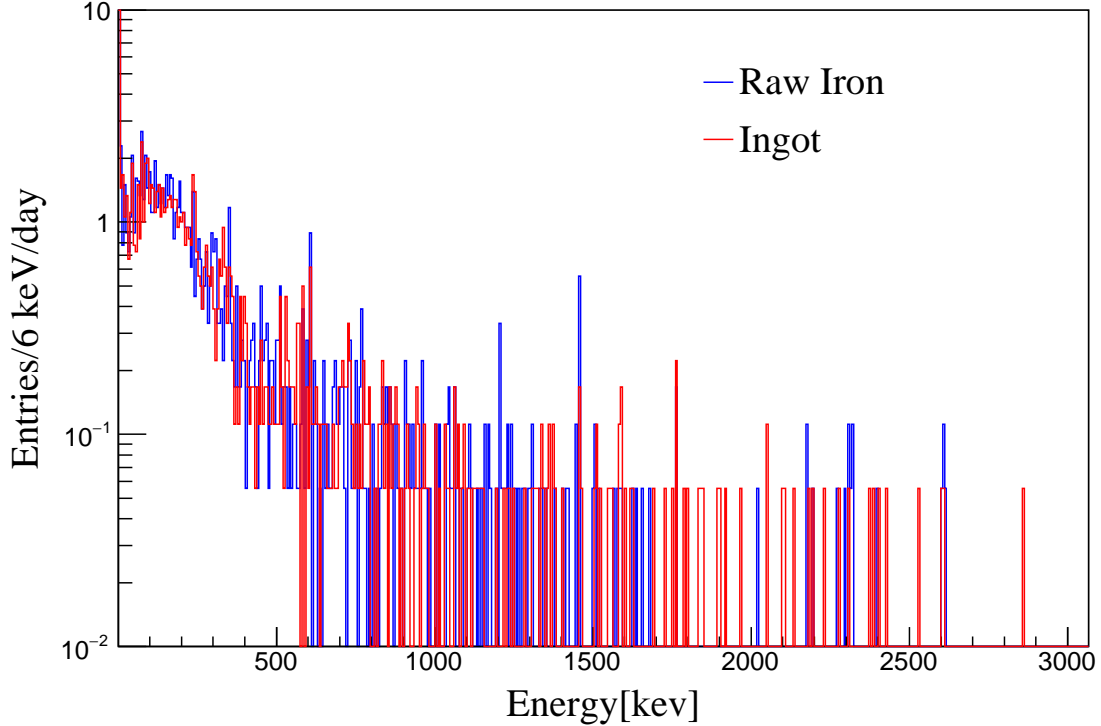}
\caption{Energy spectra of Raw Iron and Ingot measured by HPGe-Jinping.}
\label{fig:HPGe-iron}
\end{figure}

\begin{figure}
\includegraphics[width=15cm]{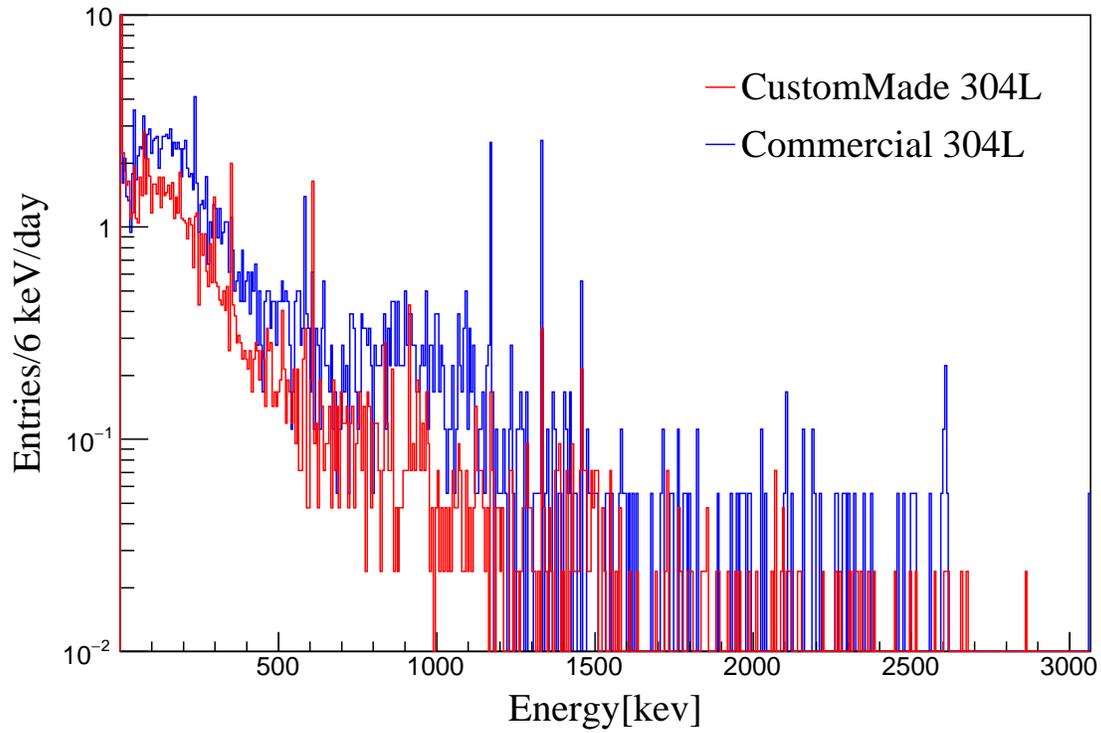}
\caption{Energy spectra of Commercial 304L and CustomMade 304L measured by HPGe-Jinping.}
\label{fig:HPGe-profile304}
\end{figure}

\begin{figure}
\includegraphics[width=15cm]{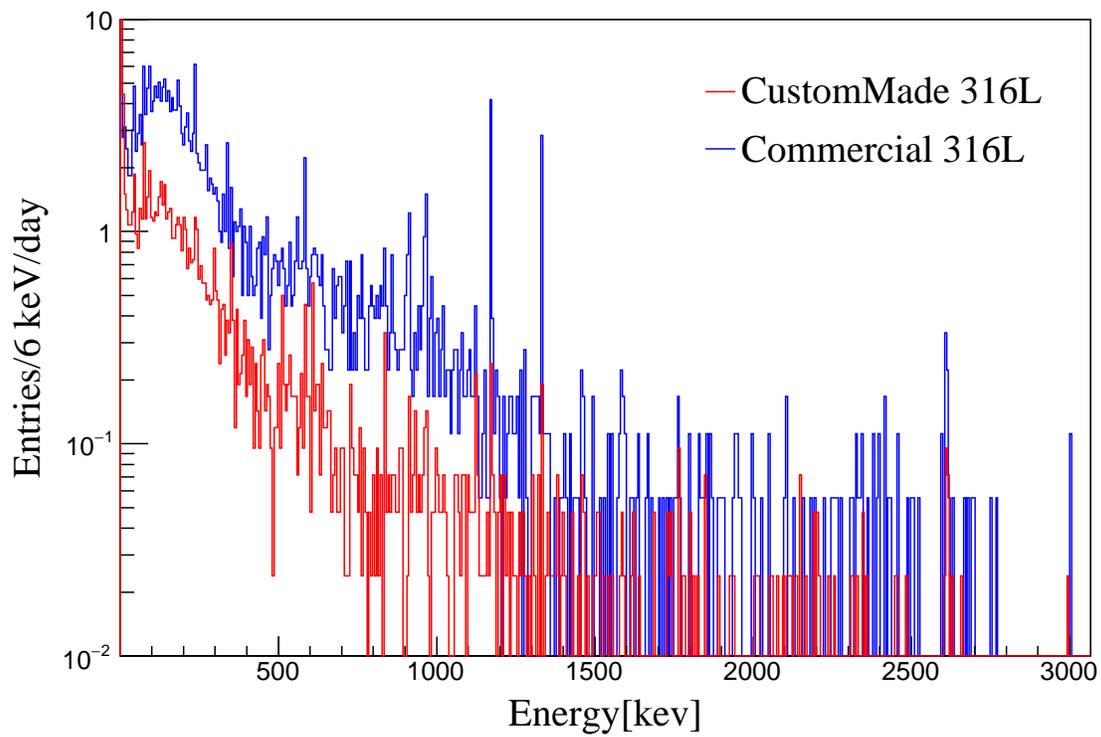}
\caption{Energy spectra of Commercial 316L and CustomMade 316L measured by HPGe-Jinping.}
\label{fig:HPGe-profile316}
\end{figure}

Next, the four stage samples and two commercial SST samples were measured with the HPGe-Jinping.
Figure~\ref{fig:HPGe-iron} shows the energy spectra and counting rates of the Raw Iron and Ingot,
Fig.~\ref{fig:HPGe-profile304} shows spectra of the CustomMade 304L and Commercial 304L samples, and Fig.~\ref{fig:HPGe-profile316} shows the spectra for CustomMade 316L and Commercial 316L samples.
The radioactivity of CustomMade samples are much lower than that of the Commercial samples.

Following the calculation method in reference~\cite{Zeng-2004}, the radioactivity of $\rm ^{226}Ra$, $\rm ^{238}U$, $\rm ^{228}Th$, $\rm ^{232}Th$, $\rm ^{40}K$, $\rm ^{60}Co$, and $\rm ^{137}Cs$ was measured from the counting rate of several characteristic gamma peaks.
Geant4~\cite{g41, g42} simulation was used to estimate the efficiencies of these characteristic peaks
and then the activities of these isotopes were calculated. The results are tabulated in table~\ref{Radio7day}.
The errors of these results are statistical uncertainty. Other uncertainties are not significant.

The radioactivity of $\rm ^{238}U$ is extracted with a joint fit for 186 keV ($^{226}$Ra), 295 keV and 352 keV ($^{214}$Pb), 609 keV, 1120 keV and 1765 keV ($^{214}$Bi), and 1001 keV ($\rm ^{234m}Pa$) assuming secular equilibrium status. For $\rm ^{226}Ra$ as a daughter nucleus of $\rm ^{238}U$, its decay chain is easy to reach secular equilibrium and its result is extracted by a joint fit without considering the 1001 keV ($\rm ^{234m}Pa$).
We also tried to extract only the radioactivity of $\rm ^{234m}Pa$ from the 1001 keV gamma, but the statistics is low.
With the current precision, the secular equilibrium assumption is valid.

The radioactivity of $\rm ^{232}Th$ is extracted with a joint fit for 209 keV, 338 keV, 911 keV, and 969 keV ($^{228}$Ac), and 583 keV and 2614 keV ($^{208}$Tl) assuming secular equilibrium. For $\rm ^{228}Th$ as a daughter nucleus of $\rm ^{232}Th$, its decay chain is easy to reach secular equilibrium and its result is extracted by a joint fit without considering all the peaks from $^{228}$Ac.
The signal strength of both the $\rm ^{228}Th$ chain and the rest of $\rm ^{232}Th$ chain is significant.
With the current precision, the secular equilibrium assumption is valid.

The radioactivity of $\rm ^{40}K$ is extracted with 1460 keV line, $\rm ^{60}Co$ with 1173 keV and 1332 keV lines, and $\rm ^{137}Cs$ with 662 keV line.

The CustomMade samples have a much cleaner background situation for $\rm ^{228}Th$, $\rm ^{232}Th$, $\rm ^{40}K$, and $\rm ^{60}Co$ than the Commercial samples. The overall counting rates of CustomMade samples are a few times smaller also.
The radioactivity of $\rm ^{226}Ra$ and $\rm ^{238}U$ is low in almost all samples but higher in CustomMade 304L.
Concerning the highest $\rm ^{40}K$ activity in Raw Iron, could it be related to the cleaning/handling of samples.
Currently no clear explanation can be found.


The GDMS and HPGe-Jinping results are compared.
We observed that the Commercial samples are more contaminated than the CustomMade samples in many elements, like Pb, Bi, Co, P, Si and Mn in GDMS results as seen in Fig.~\ref{fig:GDMS-profile}.
The GDMS results can be used as a fast indicator for cleanliness.
Note that the GDMS results are semiquantitative, so that the agreement of each value is not exact.

The detection limit of GDMS for U and Th element is $5\times 10^{-9}$ g/g, which corresponds to 62 mBq/kg for $\rm ^{238}U$, 20 mBq/kg for $\rm ^{232}Th$, the GDMS sensitivity are too low to be compared with HPGe-Jinping results.

For K element concentration, GDMS analysis got about 0.02 - 0.06 mg/kg in Commercial and CustomMade samples (an exception is found in the Ingot sample). The corresponding value for $\rm ^{40}K$ is 0.6 - 1.9 mBq/kg taking the natural abundance 0.012\% as input. The HPGe-Jinping results are about 10 times of these.

For Co element concentration, GDMS analysis got 742 mg/kg in the Commercial sample and about 12 mg/kg in the CustomMade samples.
HPGe-Jinping found positive signals of $\rm ^{60}Co$.
The HPGe-Jinping results of $\rm ^{60}Co$ in Commercial samples are also about 20 times higher than CustomMade samples, for this point, it agrees with GDMS conclusion.

For Cs element, no positive signals are found with GDMS or HPGe.

\begin{sidewaystable}
\footnotesize
\begin{tabular*}{220mm}{@{\extracolsep{\fill}}ccccccccccccc}
\toprule
  Sample					&Technique		&$\rm ^{238}U$	&$\rm^{226}Ra$	&$\rm^{234m}Pa$ &$\rm ^{232}Th$	&$\rm^{228}Th$	 &$\rm^{40}K$	&$\rm ^{60}Co$	&$\rm ^{137}Cs$	\\
  							&				&mBq/kg			&mBq/kg			&mBq/kg		    &mBq/kg			&mBq/kg			 &mBq/kg		    &mBq/kg			&mBq/kg\\
  \hline
   Raw Iron				 	&HPGe-Jinping	&9.6~$\pm$~2.2	&9.5~$\pm$~2.2	&$<$227 &2.7~$\pm$~1.2	&2.4~$\pm$~1.1	&53~$\pm$~25	 &$<$~3.8		&$<$~13\\
   Ingot					&HPGe-Jinping	&2.6~$\pm$~0.8	&2.5~$\pm$~0.8	&$<$98 &0.9~$\pm$~0.4	&0.8~$\pm$~0.4	 &8.0~$\pm$~5.7	&$<$~0.5		&$<$~2.3 \\

  CustomMade 304L			&HPGe-Jinping	&16.2~$\pm$~1.8	&16.0~$\pm$~1.8	&$<$177 &1.8~$\pm$~0.7	&1.5~$\pm$~1.1	&25~$\pm$~11	 &2.4~$\pm$~0.8	&$<$~1.2 		\\
  CustomMade 316L			&HPGe-Jinping	&6.8~$\pm$~1.2	&6.8~$\pm$~1.2	&$<$144 &4.6~$\pm$~1.5	&4.6~$\pm$~1.7	 &10.9~$\pm$~7.2	&2.0~$\pm$~0.7	&$<$~1.2 		\\

  Commercial 304L			&HPGe-Jinping	&5.2~$\pm$~1.8	&5.1~$\pm$~1.8	&$<$541 &7.4~$\pm$~2.0	&7.1~$\pm$~2.1	&49~$\pm$~23	 &22.3~$\pm$~3.4	&$<$~10 		\\
 Commercial 316L			    &HPGe-Jinping	&5.3~$\pm$~2.1	&5.2~$\pm$~2.1	&$<$484 &14.4~$\pm$~2.9	&9.6~$\pm$~2.4	&34~$\pm$~22	 &35.2~$\pm$~4.5	&$<$~12 		\\
  \hline
\end{tabular*}
\caption{\label{Radio7day}
Comparison of the radioactivity levels of six samples: Raw Iron, Ingot, CustomMade 304L, CustomMade 316L, Commercial 304L, and Commercial 316L. The radioactivity levels are given in mBq/kg, and upper limits are given at 95\% C.L.}
\end{sidewaystable}

\section{Comparison with other experiments}\label{CR}
The results are compared in table~\ref{tab:jpbkg} with those from the Borexino, GERDA, NEXT, PandaX, XENON experiments, and one of the best samples for SNOLAB.
HPGe-Jinping $\rm ^{40}K$ result indicates that it needs to be controlled for the next SST production for detectors.
There is some distance to the result of SST samples used by GERDA, NEXT, PandaX, and XENON experiments.
The activities of $\rm ^{238}U$ (or $\rm ^{226}Ra$), $\rm ^{232}Th$ (or $\rm ^{228}Th$), and $\rm ^{60}Co$ are more comparable or closer to the others.
We conclude that the SST made by smelting can be hopefully used for future experiments that require ultra-low radioactive background.

\begin{sidewaystable}
\footnotesize
\begin{tabular*}{220mm}{@{\extracolsep{\fill}}cccccccccc}

\hline\hline
Experiment			&Sample					&Supplier			&$\rm ^{238}U$		&$\rm^{226}Ra$	&$\rm ^{232}Th$	 &$\rm^{228}Th$	&$\rm^{40}K$	&$\rm ^{60}Co$\\ 						
                    &				&					&mBq/kg			&mBq/kg			&mBq/kg			&mBq/kg			&mBq/kg		 &mBq/kg\\ \hline
Jinping		& CustomMade 304L& CISRIG		&16.2~$\pm$~1.8	&16.0~$\pm$~1.8	&1.8~$\pm$~0.7	&1.5~$\pm$~1.1	&25~$\pm$~11	 &2.4~$\pm$~0.8	\\
Jinping		& CustomMade 316L& CISRIG		&6.8~$\pm$~1.2	&6.8~$\pm$~1.2	&4.6~$\pm$~1.5	&4.6~$\pm$~1.7	&10.9~$\pm$~7.2	 &2.0~$\pm$~0.7	\\

Borexino~\cite{Borexino-2002}	& AISI304L (SST for sphere)&		& - 	&4.6~$\pm$~0.9	& -	&11.4~$\pm$~1.1	&$<$~14		 &6~$\pm$~1\\

GERDA~\cite{Gerda-2008}	& 1.4571 (G5)	&Nironit (Acroni)		& -		&1.0~$\pm$~0.6	&-	&1.5~$\pm$~0.2	 &$<$~0.81		 &18.3~$\pm$~0.7\\

GERDA~\cite{Gerda-2008}	& 1.4571 (G2)	&Nironit (Ilsenburg)	& -		&$<$0.24	    &-	&$<$~0.11        &$<$0.93		 &14.0~$\pm$~0.1\\

NEXT~\cite{Next-2013}		& 304L		&Pfeiffer		& -				&14.3~$\pm$~2.8	&9.7~$\pm$~2.3	&16.2~$\pm$~3.9	&$<$~17			 &11.3~$\pm$~2.7\\
NEXT~\cite{Next-2013}		& 316Ti (10-mm-thick)&Nironit  	&$<$~21			&$<$~0.57		&$<$~0.59		&$<$~0.54		 &$<$~0.96		&2.8~$\pm$~0.2\\

PandaX-II~\cite{PandaX-2016}	&304L P03	&TISCO	&-			&1.3~$\pm$~1.1 	&- 			&6.1~$\pm$~1.9	&$<$~8.1			 &1.0~$\pm$~0.5\\
PandaX-I~\cite{PandaX-2016}		&316Ti &Changrong	&-			&$<$~1.7 		&- 	&$<$~2.2			&$<$~12.8		 &5.6~$\pm$~0.8\\

SNOLAB~\cite{SNOLAB}		&316L (Sample A)     &USA		&-	&0.63~$\pm$~0.15	      & 0.62~$\pm$~0.15	&-				& $<$~0.97		& $<$~0.15\\

XENON1T~\cite{XENON1T}  	&ANSI 304L (Sample 9) & Nironit & $<$40		        & $<$0.64 & $<$0.81	        & $<$0.36	    & $<$2.7	& 9.7~$\pm$~0.8 \\
 \hline
\end{tabular*}
\caption{Comparison of the radioactivity of the Jinping CustomMade samples and the SST samples from different experiments. Some values are not available because they are not directly given in the references.}
\label{tab:jpbkg}
\end{sidewaystable}

\section{Conclusion}\label{CONC}
Stainless steel is an important material used in neutrino experiments, and will be utilized in our upcoming rare neutrino event experiments. As such, it is crucial to understand the associated radioactive contamination.

We thoroughly investigated stainless steel samples produced by smelting and included  all raw materials, stage samples, and commercial samples in our study.
The GDMS, HPGe-Ground, and HPGe-Jinping are used to measure the contaminations in these samples.

The GDMS method is convenient and efficient and, based on table~\ref{tab:composition} and Fig.~\ref{fig:GDMS-profile}, the impurity of SST
is easily observed. Although, this method cannot directly provide results for radioactive isotopes,
it provides a strong indication when a clean SST sample is obtained.

The HPGe-Ground's sensitivity is low. However, being at ground level, it is much easier to access.
In the initial stage of this study, this facility is used to screen several raw materials and quickly identified a few potential contamination sources. Specifically, the MgO-sand, C, and Si samples are identified as active, so the HPGe-Ground results helped us to decide the next assay approach with HPGe-Jinping.

HPGe-Jinping has been installed and is currently operating under  stable conditions at the China Jinping Underground Laboratory and it is used in the sample screening of various experiments. This detector is very sensitive in the measurement of low-background materials and it provided the final results of the CustomMade samples.

The results showed that the CustomMade SST samples to have a comparable radioactivity to those used in the several other experiments. As such, stainless steel produced by this smelting process can be hopefully used in future experiments.

\section{Acknowledgements}
This work is supported in part by, the National Natural Science Foundation of China (Nos. 11235006, 11475093 and 11620101004), the Key Laboratory of Particle \& Radiation Imaging (Tsinghua University), and the CAS Center for Excellence in Particle Physics (CCEPP). We'd like to thank the PandaX collaboration in sharing the experience about low background stainless steel.

\begin{appendices}
\section{Appendix A}\label{App}
The detailed results of the GDMS analysis corresponding to each element of Raw Iron, Ingot, Commercial 316L, CustomMade 304L, and CustomMade 316L are presented in table ~\ref{tab:long}.

\begin{center}
\footnotesize
\begin{longtable}{cccccc}
\caption{Results of GDMS analysis for elements Li to Es.} \label{tab:long} \\

\hline\hline \multicolumn{1}{c}{\textbf{Element}} & \multicolumn{1}{c}{\textbf{Raw Iron}} & \multicolumn{1}{c}{\textbf{Ingot}} & \multicolumn{1}{c}{\textbf{Commercial 316L}} & \multicolumn{1}{c}{\textbf{CustomMade 304L}} & \multicolumn{1}{c}{\textbf{CustomMade 316L}}\\
\multicolumn{1}{c}{\textbf{}} & \multicolumn{1}{c}{\textbf{mg/kg}} & \multicolumn{1}{c}{\textbf{mg/kg}} & \multicolumn{1}{c}{\textbf{mg/kg}} & \multicolumn{1}{c}{\textbf{mg/kg}} & \multicolumn{1}{c}{\textbf{mg/kg}}\\ \hline
\endfirsthead

\multicolumn{6}{c}%
{{\bfseries \tablename\ \thetable{} -- continued from previous page}} \\
\hline \hline  \multicolumn{1}{c}{\textbf{Element}} & \multicolumn{1}{c}{\textbf{Raw Iron}} & \multicolumn{1}{c}{\textbf{Ingot}} & \multicolumn{1}{c}{\textbf{Commercial 316L}} & \multicolumn{1}{c}{\textbf{CustomMade 304L}} & \multicolumn{1}{c}{\textbf{CustomMade 316L}}\\
\multicolumn{1}{c}{\textbf{}} & \multicolumn{1}{c}{\textbf{mg/kg}} & \multicolumn{1}{c}{\textbf{mg/kg}} & \multicolumn{1}{c}{\textbf{mg/kg}} & \multicolumn{1}{c}{\textbf{mg/kg}} & \multicolumn{1}{c}{\textbf{mg/kg}}\\ \hline
\endhead

\hline \multicolumn{6}{r}{{Continued on next page}} \\
\endfoot
\hline
\endlastfoot
        Li	&$<$0.005       	&$<$0.005	&$<$0.005               	&$<$0.005               	&$<$0.005\\
        Be  	&$<$0.005       	&$<$0.005 	&$<$0.005                &$<$0.005               	&$<$0.005\\
        B   	&0.11              	&0.53                &3.25                        & 3.99                      	&4.02\\
        F   	&0.035            	&0.088              &0.1                         	&0.028                      &0.013\\
        Na  	&0.051            	&0.11                & 0.007                     &0.025                      &0.014\\
        Mg  	&0.009             	&0.077              &0.25                 	&0.61                        &2.17\\
        Al  	&26.8              	&0.050              &1.79                        &1.62                        &7.81\\
        Si  	&43.9              	&8.77                &1671                       &261                         &284\\
        P   	&40.2              	&19.9                &118                         &14.6                        &15.4\\
        S   	&11                	&14.3                &$<$0.5                	&8.58                        &8.25\\
    	Cl 	&0.23        	&0.74                &$<$0.5             	&0.36            		&0.29\\
    	K   	&0.019        	&0.20                &0.02               		&0.057            		&0.033\\
    	Ca  	&$<$0.005  	&$<$0.005        &0.21           		&$<$0.005        	&0.050\\
    	Sc  	&$<$0.005       &$<$0.005         &$<$0.5          		&$<$0.005        	&$<$0.005\\
    	Ti  	&0.21        	&0.019              &0.49        		&3.38            		&8.36\\
    	V  	&1.05        	&1.12                &197            		&75.4            		&123\\
    	
	Cr    &137        		&Contained                    	&Contained           &Contained            &Contained\\
    	Mn   &145        		&61.1                    		&3853            	     &233                &202\\
    	Fe    &Contained			&Contained			&Contained	&Contained			&Contained \\
	Co   &12.6    		&11.6                    		&742                		&13.3            &8.39\\
    	Ni    &40.8    		&Contained                    	&Contained           		 &Contained           &Contained\\
	Cu  &28.4              & 34.9                                   &1188                           & 16.1              &14.8\\
	 Zn  & 0.18             &0.072                                   &3.99                           & 0.072              &0.11\\
        Ga  &8.05              &3.13                                    &11.1                           & 6.65              &8.85\\
        Ge  &9.69              &1.59                                    & 2.87                          & 1.82              &2.33\\
	As  & 3.01             &6.98                                    &17.1                           & 6.07              &6.84\\
        Se  &$<$0.005          &$<$0.005                         & 0.025                & 0.26              &0.24\\
	Br  	&0.91              	&0.012              &$<$0.005               	&0.008               	&0.006\\
        Rb  	&$<$0.005       &$<$0.005         &$<$0.005              	&$<$0.005        	&$<$0.005\\
        Sr  	&$<$0.005       &$<$0.005         &0.043  		       	&0.034              	&0.048\\
        Y   	&$<$0.005   	&$<$0.005        &0.005                  	&0.009              	&0.021\\
 	Zr  	&0.18		&0.011		&1.2                          &0.35              		&0.49\\
        Nb  	&0.051           	&0.27		& 8.8                         &10.8              		&13.5\\
        Mo  	&4.23           	&2.60		&1530                       &48              		&83.0\\
                Tc     &-            		&$<$0.005                        &-                            &$<$0.005               &$<$0.005\\

        Ru  	&$<$0.005       &$<$0.005		&0.086                 	&0.007              	&$<$0.005\\
        Rh  	&0.008             &$<$0.005        &0.023                 	&$<$0.005              	&$<$0.005\\
        Pd  	&$<$0.005       &$<$0.005        &0.053                 	&0.038             		&$<$0.005\\
        Ag  	&$<$0.005  	&0.016             &0.3                           &0.020             		&0.006\\
        Cd  	&$<$0.005 	&0.021             &0.25                         &0.016            		&0.76\\
	In  	&0.28             	&$<$0.005                      	&0.069                  &0.005             	&$<$0.005\\
        Sn  	&0.55            	&0.14                              	&16.05                  &0.20              	&0.31\\
        Sb  	&2.33            	&0.49                              	&3.7                      &0.25             	&0.35\\
        Te  	&$<$0.005        &0.009                           	&$<$0.005            &$<$0.005            &$<$0.005\\
        I   	&$<$0.005       	&$<$0.005                     	&$<$0.005            &$<$0.005            &$<$0.005\\
        Cs  	&$<$0.005      	&$<$0.005                     	&$<$0.005            &$<$0.005          	&$<$0.005\\
        Ba  	&$<$0.005        &$<$0.005                    	&0.46                    &$<$0.005            &0.14\\
        La  & $<$0.005         &$<$0.005                         &$<$0.005               &$<$0.005               &$<$0.005\\
        Ce  &$<$0.005          &$<$0.005                         & 0.2                         & $<$0.005              &0.070\\
        Pr  & $<$0.005         &$<$0.005                                & $<$0.005              & $<$0.005              &$<$0.005\\
	Nd  &$<$0.005          &0.009                            & $<$0.005              &$<$0.005               &$<$0.005\\
        Pm   &-             		&$<$0.005                        &-                            &$<$0.005               &$<$0.005 \\

        Sm  &$<$0.005          &$<$0.005                         & $<$0.005              &$<$0.005               &$<$0.005\\
        Eu  &$<$0.005          &$<$0.005                         & $<$0.005              &$<$0.005               &$<$0.005\\
        Gd  & $<$0.005         &$<$0.005                         & $<$0.005              & $<$0.005              &$<$0.005\\
        Tb  &$<$0.005          &$<$0.005                         & $<$0.005              & $<$0.005              &$<$0.005\\
        Dy  &$<$0.005          &$<$0.005                         & $<$0.005              &$<$0.005               &$<$0.005\\
        Ho  &$<$0.005          &$<$0.005                         & $<$0.005              &$<$0.005               &$<$0.005\\
  	Er  &$<$0.005          &$<$0.005                         & $<$0.005              &$<$0.005               &$<$0.005\\
        Lu         &-             &$<$0.005                         &-                                  &$<$0.005               &$<$0.005\\	
        Tm   	&$<$0.005        &$<$0.005                    	&$<$0.005            &$<$0.005            &$<$0.005\\
        Yb   	&$<$0.005        &$<$0.005                    	&$<$0.005            & $<$0.005           &$<$0.005\\
        Lu   	&$<$0.005        &$<$0.005                    	&$<$0.005            &$<$0.005            &$<$0.005\\

        Hf   	&$<$0.005        &$<$0.005                    	&$<$0.005            & $<$0.005           &$<$0.005\\
        Ta     &0.042        	&0.035                               &0.049                  &0.11               	&0.18  \\
        W     &0.58          	&0.068                               &58                       &0.095               	&0.99  \\
        Re    &0.028         	&0.006                               &0.28                    & $<$0.005           &0.005  \\
        Os   	&$<$0.005        &$<$0.005                        &0.024                   &$<$0.005            &$<$0.005\\
        Ir   	&$<$0.005        &$<$0.005                        &0.026                   &$<$0.005            &$<$0.005\\
        Pt   	&$<$0.005        &$<$0.005                        &0.095                   &$<$0.005            &$<$0.005\\
        Au  	&0.035              &$<$0.005                        &0.093                   & 0.019              	&0.013\\
        Hg    &0.28         	&$<$0.005                        &$<$0.005             &$<$0.005               &0.006\\
        Tl      &$<$0.005     	&$<$0.005                        &$<$0.005             &$<$0.005               &$<$0.005\\
        	 Pb     &$<$0.005     	&$<$0.005                      	&0.074                   &$<$0.005               &0.009\\
        Bi      &$<$0.005     	&$<$0.005                       	&0.03                     &$<$0.005               &$<$0.005\\
        	Po         &-             &$<$0.005                         &-                                  &$<$0.005               &$<$0.005\\
        Th     &$<$0.005     	&$<$0.005                    	& $<$0.005            &$<$0.005               &$<$0.005\\
       Pa         &-             &$<$0.005                         &-                                  &$<$0.005               &$<$0.005\\
        U      &$<$0.005     	&$<$0.005                   	& $<$0.005            &$<$0.005               &$<$0.005\\
        Np         &-             &$<$0.005                         &-                                  &$<$0.005               &$<$0.005\\
	Pu         &-             &$<$0.005                         &-                                  &$<$0.005               &$<$0.005\\
        Am         &-             &$<$0.005                        &-                                   &$<$0.005               &$<$0.005\\
        Cm         &-             &$<$0.005                	    &-                                   &$<$0.005               &$<$0.005\\
        Bk         & -            &$<$0.005                         &-                                   &$<$0.005               &$<$0.005\\
        Cf         &-             &$<$0.005                         &-                                    & $<$0.005              &$<$0.005\\
        Es         &-             &$<$0.005                        &-                                    &$<$0.005               &$<$0.005\\
\end{longtable}
\end{center}
\end{appendices}
\end{document}